# Antiferro-quadrupole resonance in CeB$_6$

S.V. Demishev[a], A.V. Semeno[a], A.V. Bogach[a], Yu.B. Paderno[b], N.Yu. Shitsevalova[b] and N.E. Sluchanko[a]

[a] *A.M.Prokhorov General Physics Institute of Russian Academy of Sciences, Vavilov street, 38, 119991 Moscow, Russia*

[b] *Institute for Problems of Materials Science of Ukranian National Academy of Sciences, Krzhyzhanovsky street, 3, 03680 Kiev, Ukraine*

**Abstract**

We report experimental observation of a new type of magnetic resonance caused by orbital ordering in a strongly correlated electronic system. Cavity measurements performed on CeB$_6$ single crystals in a frequency range 60-100 GHz show that a crossing of the phase boundary $T_Q(B)$ between the antiferro-quadrupole and paramagnetic phases gives rise to development at $T<T_Q(B)$ of a magnetic resonance. The observed mode is gapless and correspond to g-factor 1.62.

Cerium hexaboride is a well-known example of a compound where quadrupole magnetic interactions play an essential role [1-6]. The dipole magnetic and quadrupole electric magnetic moments are believed to be associated with the Ce$^{3+}$ ions forming a simple cubic lattice [1,2]. In zero magnetic field the quadrupole (i.e. orbital) ordering with formation of an antiferro-quadrupole (AFQ) phase occurs at $T_Q$=3.2 K and precedes the formation of an antiferromagnetic (AF) phase (i.e. dipole ordering) at $T_D$=2.3 K. In the region $T>T_Q$ cerium hexaboride is a paramagnetic metal (P phase) and demonstrates behaviour typical of a dense Kondo-system. The application of the external magnetic field induces an enhancement of $T_Q$ and suppression of $T_D$ [1-6]

The coupling between electric quadrupole moments and dipole magnetic moments suggests the study of the antiferro-quadrupole phase of CeB$_6$ by means of various magnetic techniques, not excluding *a priori* resonant measurements. However, for dense Kondo-systems the spin fluctuations at the magnetic ions are generally believed to broaden the resonance line width so much that the magnetic resonance becomes undetectable [7]. The Kondo temperature for CeB$_6$ is $T_K$~1 K [1-6] and the expected line width will be about [7] $W \sim k_B T_K/\mu_B \sim 1.5$ T. Therefore EPR-like modes with a g-factor value close to 2 may be observed in CeB$_6$ at frequencies about 60 GHz, corresponding to a resonant field of about 2 T and in the temperature range $T>T_K$~1 K including the AFQ phase.

In our experiment we have measured the transmission of the copper cylindrical cavities operating at TE$_{01n}$ modes and tuned to frequencies 60, 78 and 100 GHz. One of the endplates of the cylinder has been made of the high quality CeB$_6$ single crystal. The quality factor was about ~6·10$^3$. The cavity has been placed in the cryostat with a 7 T superconducting magnet and the experimental setup has allowed controlling the cavity temperature with an accuracy better than 0.01 K down to 1.7 K. The magnetic field was aligned along the [110] crystallographic direction and was parallel to the cavity axis. As a reference a small DPPH sample has been placed in the cavity. Detais about samples quality control and characterisation can be found elsewhere [8].

The obtained data are summarised in fig. 1. For $T>T_Q(B)$ the field dependence of the cavity transmission displays a bend, which shifts to lower magnetic field when temperature is lowered (fig. 1). This feature reflects the change of the CeB$_6$ impedance at the boundary between paramagnetic and antiferro-quadrupole phases.

In contrast to the paramagnetic phase, for $T<T_Q$ a new strong resonance develops and the intensity of this line increases with lowering temperature. Frequency measurements have shown that the position of the observed line shifts linearly corresponding to a Lande g-factor 1.62, which strongly deviates from the values $g$=1.98-2.5 reported for the EPR in the paramagnetic phase [9]. At low temperatures the bend of the transmission curves corresponding to the boundary between antiferromagnetic and antiferro-quadrupole phase has been also detected (fig. 1).

Note that the positions of the bends in fig. 1 completely agree with magnetic phase diagram subtracted from the magnetization and magnetoresistance experiments (fig. 2).

In described experiment both bulk and surface properties of the sample may contribute to a cavity microwave response. Therefore various kinds of the sample surface preparations have been checked, including several roughnesses of the mechanical polishing and different regimes of chemical etching. Experiments with various surface treatments have provided results identical to those presented above; the excitation of the resonant mode below $T_Q$ has been also checked for several samples. Note that observation of the phase boundaries subtracted from the volume properties like magnetisation [4-5] and resistivity [6] in the same cavity experiment also favours the understanding of the observed resonance as a bulk effect.

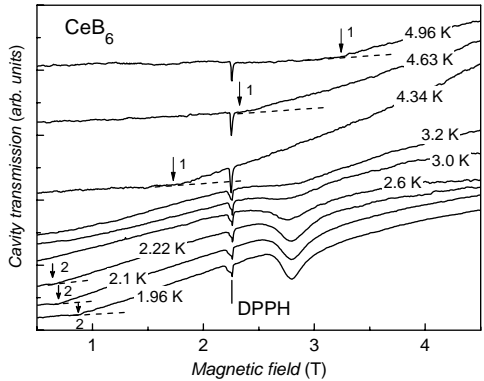

Fig. 1. Field dependences of cavity transmission at various temperatures. Arrows mark magnetic phase transitions: (1) AFQ phase – P phase); (2) AFphase - AFQ phase.

We have analysed the temperature dependences of the line width $W(T)$ and integrated intensity $I(T)$ [8]. It is found that approaching of the $T_Q(B_{res})$ from below causes a divergence line width accompanied by a strong decrease of the integrated intensity. Namely in the range 1.8-3.7 K the $W(T)$ increases 2.6 times whereas $I(T)$ drops by a factor of 3 thus making the resonance undetectable above 3.7 K (see fig. 1 and Ref. 8).

Summarising, we have observed a novel magnetic resonance specific to the antiferro-quadrupole phase of $CeB_6$. From the data obtained it is natural to suppose that the properties of the localized magnetic moments at the antiferro-quadrupole boundary in $CeB_6$ should change dramatically making magnetic resonance mode possible. Unfrotunately, to our best knowledge, no theory relevant to magnetic resonances in the antiferro-quadrupole phase exists up to now.

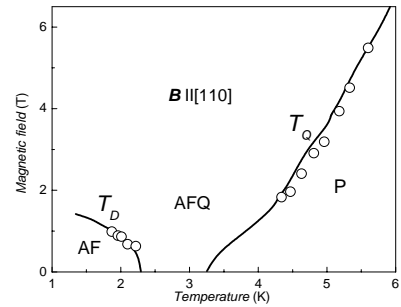

Fig. 2 Magnetic phase diagram of $CeB_6$ in cavity experiment (points, present work) and in magnetisation and resistivity measurements (solid lines, Ref. 4-6).

This work was supported by the INTAS project 03-51-3036 and by the program of the RAS "Strongly Correlated Electrons". Part of the work was supported by RFBR grants 04-02-16574 and 04-02-16721.